\documentclass[%
 reprint,
superscriptaddress,
 amsmath,amssymb,
 aps,
prb,
]{revtex4-2}

\usepackage{soul}
\usepackage{amssymb}
\usepackage{tabularx}
\usepackage{multirow}
\usepackage[colorlinks]{hyperref}
\usepackage{soul,color}
\usepackage{graphicx}
\usepackage{dcolumn}
\usepackage{bm}
\usepackage{subcaption}

\begin{document}

\title{The influence of constriction on the motion of graphene kinks}

\author{Dyk Chung Nguyen}
\affiliation{%
 Department of Physics and Astronomy, University of South Carolina, Columbia, South Carolina 29208, USA
}%
\author{Ruslan D. Yamaletdinov}
\affiliation{%
Nikolaev Institute of Inorganic Chemistry SB RAS,
Novosibirsk, 630090, Russia
}
\affiliation{Boreskov Institute of Catalysis SB RAS,
Novosibirsk, 630090, Russia}
\author{Yuriy V. Pershin}%
 \email{pershin@physics.sc.edu}
\affiliation{%
 Department of Physics and Astronomy, University of South Carolina, Columbia, South Carolina 29208, USA
}%

\begin{abstract}
Graphene kinks are topological states of buckled graphene membranes. We show that when a moving kink encounters a constriction, there are three general classes of behavior: reflection, trapping, and transmission. Overall, constriction is characterized by an attractive potential.
In the case of a simple symmetric constriction, the kink potential energy has a relatively deep minimum surrounded by energy barriers.
However, the potential energy alone  does not fully define the class of behavior: the effect of a resonant reflection was observed in our simulations.
Moreover, we demonstrate that asymmetric constrictions can transform kinks from one type into another. MD simulation results are compared with predictions of the classical $\phi^4$ model.
\end{abstract}

\maketitle

\section{Introduction}

Constrictions and barriers  play a tremendous role in the transport of particles, molecules, and energy. From the early days of quantum mechanics to the present, several related transport phenomena have been discovered and explored, including tunneling and resonant tunneling in quantum systems~\cite{razavy2003quantum}, conductance quantization in quantum point contacts~\cite{van1988quantized}, and the Josephson effect in tunnel junctions and weak links~\cite{likharev1979superconducting}.
In this paper we study the effect of constriction on the motion of kinks in buckled graphene~\cite{Yamaletdinov17a}. Graphene is a
$\sigma$-bond connected sheet of carbon atoms that are packed into a two-dimensional honeycomb crystal lattice. Its
outstanding properties  (such as high mechanical strength~\cite{lee2008measurement}, electrical conductivity \cite{berman2014extraordinary} and optical transparency \cite{optical_transparency,vakil2011transformation})
make graphene attractive for applications in conducting electrodes \cite{electrodes}, sensors \cite{sensors,banadaki2019graphene}, and memory devices~\cite{memristors,memcapacitors}, to name a few.
Recently, it was suggested that  the topological states of the classical $\phi^4$ theory can be implemented in buckled graphene~\cite{Yamaletdinov17a}.

We start with the introduction of the classical $\phi^4$ theory~\cite{kevrekidis2019dynamical}. It is based on the  Lagrangian density~\cite{rajaraman1975some}
\begin{equation}
    \mathcal{L} = \frac{1}{2} \left( \frac{\partial \phi}{\partial t} \right)^2 - \frac{1}{2} \left( \frac{\partial \phi}{\partial x} \right)^2 - \frac{1}{4}(1-\phi^2)^2
    \label{eq:1}
\end{equation}
depending on the field $\phi(x,t)$ and its derivatives. Eq.~(\ref{eq:1}) leads to the Euler--Lagrange equation of motion
\begin{equation}
    \frac{\partial^2 \phi}{\partial t^2} - \frac{\partial^2 \phi}{\partial x^2} + \phi^3 - \phi = 0.
    \label{eq:2}
\end{equation}
A topologically nontrivial solution for Eq.~(\ref{eq:2}) is
\begin{equation}
    \phi = \pm \tanh{\frac{x-vt-a}{\sqrt{2(1-v^2)}}},
    \label{eq:3}
\end{equation}
where the \(\pm\) sign corresponds to whether it is a kink or an antikink, $a$ is the initial position, and $v$ is the velocity. The properties of classical $\phi^4$ kinks have been extensively studied and well understood, including the features of
the dynamics of a single kink~\cite{combs1983single}, kink--antikink collisions~\cite{CAMPBELL19831,Goodman05a,Goodman07a,kink_antikink_phi4_phi6,bazeia2018scattering}, and the effect of impurities on the motion of kinks~\cite{braun1991nonlinear,ekomasov2017resonance,kink_and_realistic_impurity}.  For more on the recent developments in $\phi^4$ theory, see~\cite{kevrekidis2019dynamical}.

\begin{figure}[b]
\includegraphics[width=0.8\columnwidth]{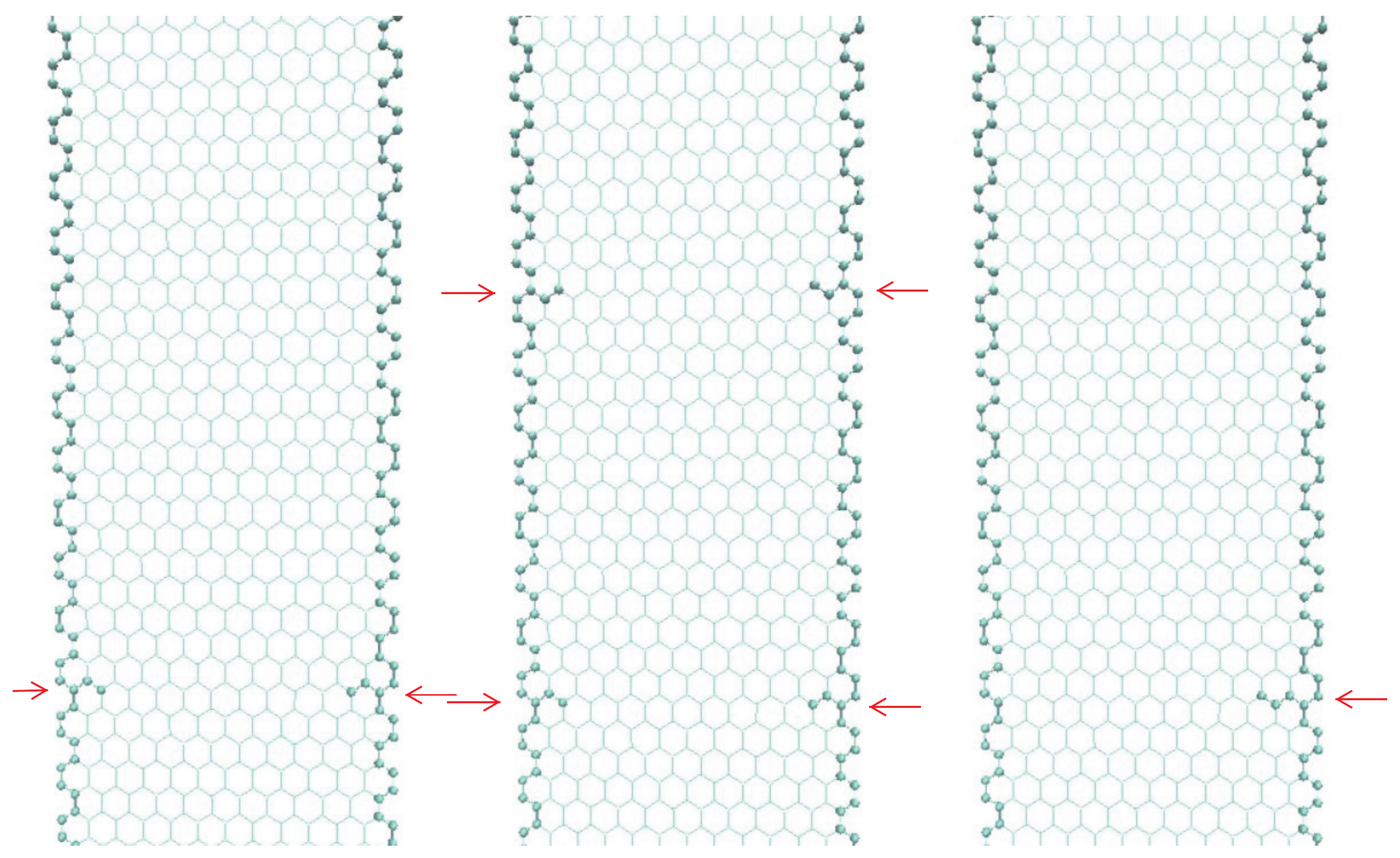}\includegraphics[width=0.058\columnwidth]{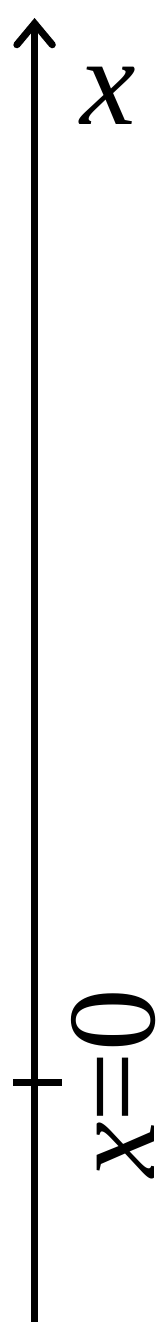}\\
Type 1 \hspace{1.2cm}Type 2 \hspace{1.2cm}Type 3
\caption{\label{fig:1}Types of constrictions considered in this work. The vertical direction is $x$, and the horizontal is $y$. The fixed atoms are represented by spheres. The arrows indicate the location of constriction atoms.}
\end{figure}

Graphene kinks and antikinks are topological states of long buckled graphene membranes with fixed long edges~\cite{Yamaletdinov17a,Yamaletdinov19a,Yamaletdinov20a}. When motionless, these states can be defined as the lowest energy conformations in which the membrane is buckled homogeneously in opposite directions in the vicinity of short opposite edges. There are many similarities between graphene kinks and $\phi^4$ kinks, including their shape and certain dynamical characteristics
such as the constant velocity motion (to a good approximation) and the relativistic-type dependence of the kinetic energy on the velocity~\cite{Yamaletdinov20a}.
At the same time, there are some differences: there are four types of graphene kinks, $(\alpha,\pm)$ and $(\beta,\pm)$, the dissipations are more important in graphene, etc. For the classification of graphene kinks see Ref.~\cite{Yamaletdinov20a}.

The main goal of this paper is to understand the influence of constrictions on the motion of graphene kinks. For this purpose, we perform a series of molecular dynamics (MD) simulations modeling the interaction of moving kinks with constrictions. Fig.~\ref{fig:1} presents the three types of constrictions considered in this work: symmetric (Type 1), double symmetric (Type 2), and asymmetric (Type 3).

This paper is organized as follows. Sec.~\ref{sec:2} describes the details of the molecular dynamics simulations. Sec.~\ref{sec:3} describes the results of the MD simulations. For each type of constriction, the potential energy profile is calculated. It is found that a type 1 symmetric constriction can be represented by a potential well surrounded by repulsive barriers (subsec.~\ref{sec:3b}). We discovered that asymmetric constrictions can flip the kink type (subsec.~\ref{sec:3c}). Moreover, in subsec.~\ref{sec:3d}, a constriction is introduced into the classical $\phi^4$ model, and the scattering of a $\phi^4$ kink on a constriction is simulated. Sec.~\ref{sec:4} concludes.

\section{Details of MD simulations}\label{sec:2}

\begin{figure*}[tb]
\begin{subfigure}[r]{.05\textwidth}
(a) \\
\vspace{1.2cm}
(b) \\
\vspace{1.2cm}
(c) \\
\end{subfigure}
\begin{subfigure}[l]{.9\textwidth}
\includegraphics[width=0.9\textwidth]{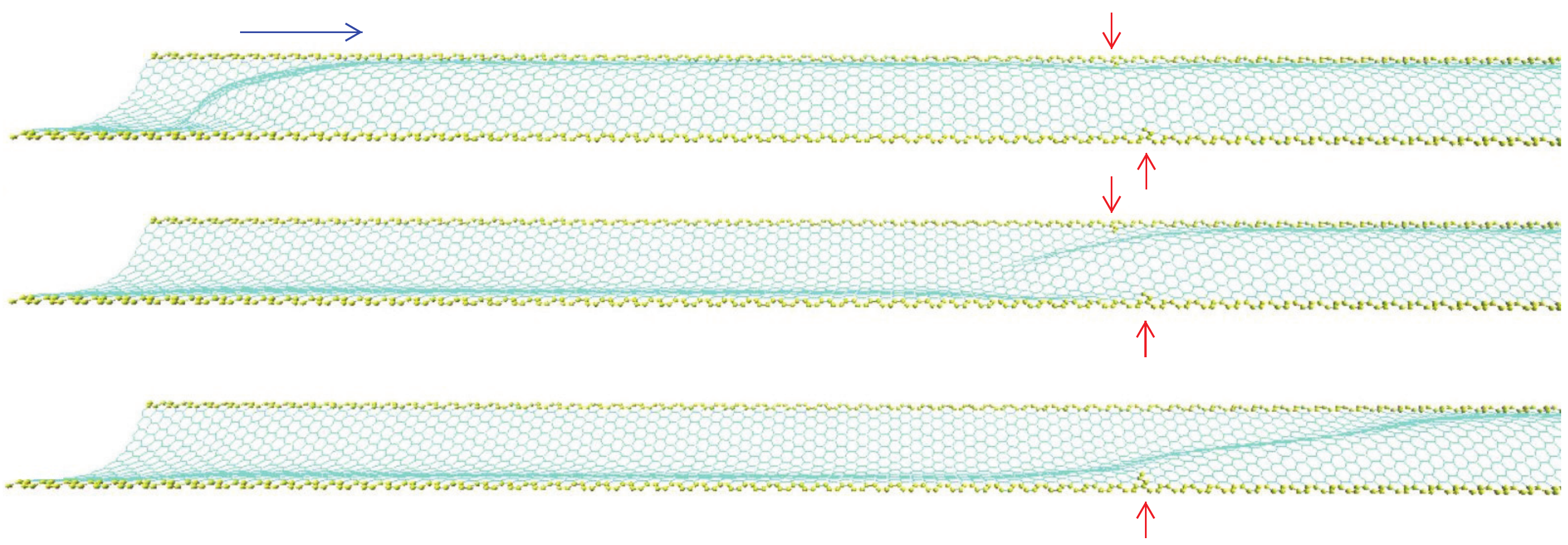}
\end{subfigure}
    \caption{\label{fig:2} Nanoribbon geometries. (a) A moving ($\alpha,+$)-kink at $t=0$. (b) An example of the final geometry with ($\alpha,+$)-kink trapped by the constriction. (c) An example of the final geometry with ($\beta,-$)-kink trapped by the constriction. The fixed atoms are shown in yellow.  The red arrows indicate the location of constriction atoms. The blue arrow represents the initial kink velocity, $v_0$.}
\end{figure*}

The MD simulations were performed using the NAMD  package~\cite{phillips2020scalable}. The atomic coordinates were visualized using VMD~\cite{HUMP96}~\footnote{NAMD and VMD were developed by the Theoretical and Computational Biophysics Group at the Beckman Institute for Advanced Science and Technology at the University of Illinois at Urbana-Champaign.}. A previously developed~\cite{Yamaletdinov17a} CHARMM type~\cite{JCC21367} force field for carbon atoms in graphene was used in our simulations. The force field includes 2-body spring bond, 3-body angular bond (including the Urey-Bradley term), 4-body torsion angle, and Lennard-Jones potential energy terms that were optimized to match the structural and physical properties of graphene such as the in-plane stiffness, bending rigidity, and equilibrium bond length~\cite{Yamaletdinov17a}.

A flat graphene nanoribbon with length 400~{\AA} and width 25~{\AA} was selected as the starting point to build the system. The membrane was compressed in the $y$-direction by scaling the $y$-coordinates of the  carbon atoms by a factor of 0.9. The first two rows of carbon atoms at the long edges were set fixed. Additionally, to create the constriction, we fixed several other atoms in the middle of the nanoribbon, as shown in Fig.~\ref{fig:1}, where the fixed atoms are drawn using the CPK method. In particular, we defined a symmetric constriction by fixing two atoms from each side (type 1), a double symmetric constriction by
placing two type 1 constrictions at a distance of
31.95~{\AA} from each other (type 2), and an asymmetric constriction by fixing three atoms from one side only (type 3).

To obtain a membrane homogeneously buckled in the $+z$-direction, we ran a series of identical simulations at $T=293$~K. The simulations were performed for 20000 steps with 1 fs per step, followed by 10000 steps of energy minimization. The Langevin damping parameter was 0.2~\(\textnormal{ps}^{-1}\). The Van der Vaals forces started decreasing at 12~{\AA} and reached 0 at 14~{\AA}.
Because of the Langevin dynamics, the evolution was different in different runs~\cite{Yamaletdinov17a}.
From these results, we selected the geometry of a membrane buckled uniformly  in the $+z$-direction.

To create a moving kink, we applied an external force $f$ in the $-z$-direction to a group of 112 atoms located in the vicinity of the shorter edge at $y=-200$~{\AA}.
This method creates moving $(\alpha,+)$-kinks when $f\gtrsim 35$~pN/atom with a minimal kink velocity of about 3~km/s~\cite{Yamaletdinov17a}.
To reduce the noise, the evolution was terminated after 3000 steps of the dynamics, and the coordinates of the atoms located outside the region of the kink were set to their equilibrium values. Moreover, their velocities were set to zero~\cite{Yamaletdinov19a}. In the subsequent simulations, this point was used as the initial condition. To create a kink moving with a specific velocity, the velocity components in an NAMD velocity file were scaled appropriately. The initial velocity of the kink was measured by dividing the displacement of the kink by the  corresponding time interval. The studies of moving kinks were performed without temperature control and Langevin damping (microcanonical ensemble).

To find the potential energy as a function of the position of the kink, we ran a series of similar kink dynamics simulations with different numbers of simulation steps (with a step size of 200~fs). Each simulation was followed by minimization. The energy and position of the kink were extracted from the  minimized geometries.

\section{Results}\label{sec:3}

Table \ref{tab:1} shows a summary of our MD simulation results. The details
are given below in subsections~\ref{sec:3a}--\ref{sec:3c}.

\begin{center}
\begin{table}
\begin{tabular}{ |c|c|c|c| }
\hline
Constriction & $v_0$,  km/s & Observations \\
\hline
\multirow{4}{4em}{Type 1} & $<$ 0.55 & Reflection from 1st barrier \\
& 0.56 - 0.81 & Trapping \\
& 0.795, 0.836 & Reflection resonances \\
& $>$ 0.84 & Mainly transmission \\
\hline
\multirow{3}{4em}{Type 2} & $<$ 0.6 & Reflection \\
& 0.6 - 1.07 & Trapping \\
& 0.934 & Transmission \\
& $>$ 1.07 & Mainly transmission \\
\hline
\multirow{8}{4em}{Type 3} & $<$ 0.5 & Reflection \\
& 0.54 & Transformation $\alpha\rightarrow\beta$, trapping \\
& 0.56 - 0.65 & Trapping \\
& 0.7 - 0.8 & Transformation $\alpha\rightarrow\beta$, trapping \\
& 0.8 - 1.7 &  $\alpha\rightarrow\beta\rightarrow\alpha$, trapping  \\
& 1.75 - 1.8 &  $\alpha\rightarrow\beta$, transmission \\
& 1.8 - 2.3 &   $\alpha\rightarrow\beta\rightarrow\alpha\rightarrow\beta$, trapping \\
& $>$ 2.3 & Transmission \\
\hline
\end{tabular}
\caption{\label{tab:1}Summary of our main results on the collision of an $(\alpha,+)$-kink with a constriction. Here, $v_0$ is the initial velocity of the kink.}
\end{table}
\end{center}

\subsection{Geometries} \label{sec:3a}

Fig.~\ref{fig:2} gives examples of the geometries of the MD simulations. Here, the membrane is buckled in the transverse direction, and stays in the buckled state supported by the fixed atoms along the longer edges.  Fig.~\ref{fig:2}(a) shows a moving ($\alpha,+$)-kink that was created as  described in Sec.~\ref{sec:2}. A geometry similar to Fig.~\ref{fig:2}(a) was used as the initial condition in our main simulations. Figs.~\ref{fig:2}(b) and (c) provide examples of the final geometries where an ($\alpha,+$)-kink is trapped by a type 1 symmetric constriction (see Fig.~\ref{fig:2}(b)), and a ($\beta,-$)-kink is trapped by a type 3 asymmetric constriction (see Fig.~\ref{fig:2}(c)).

\begin{figure}[tb]
    \centering \includegraphics[width=0.9\columnwidth]{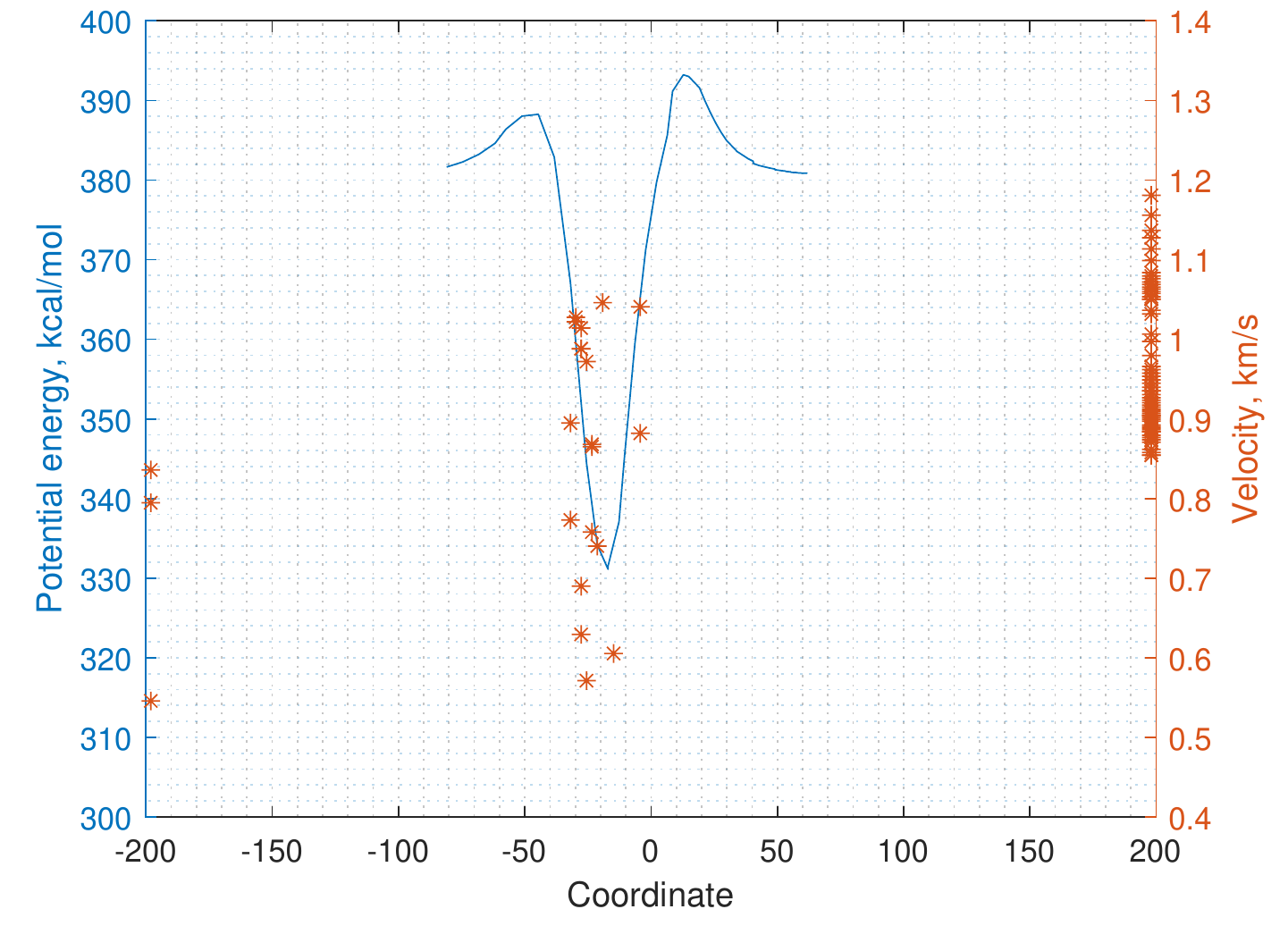}
    \includegraphics[width=0.49\columnwidth]{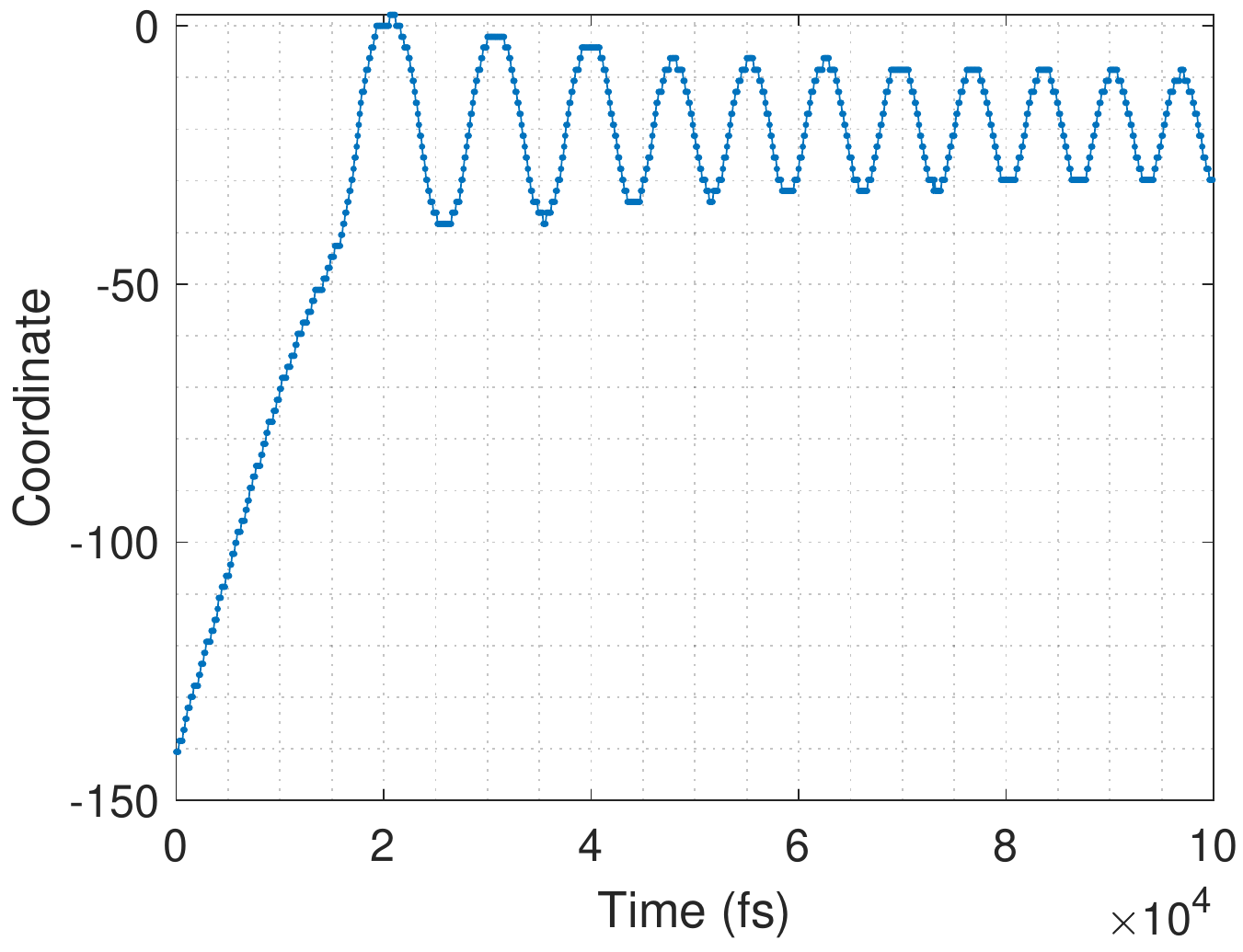}
    \includegraphics[width=0.49\columnwidth]{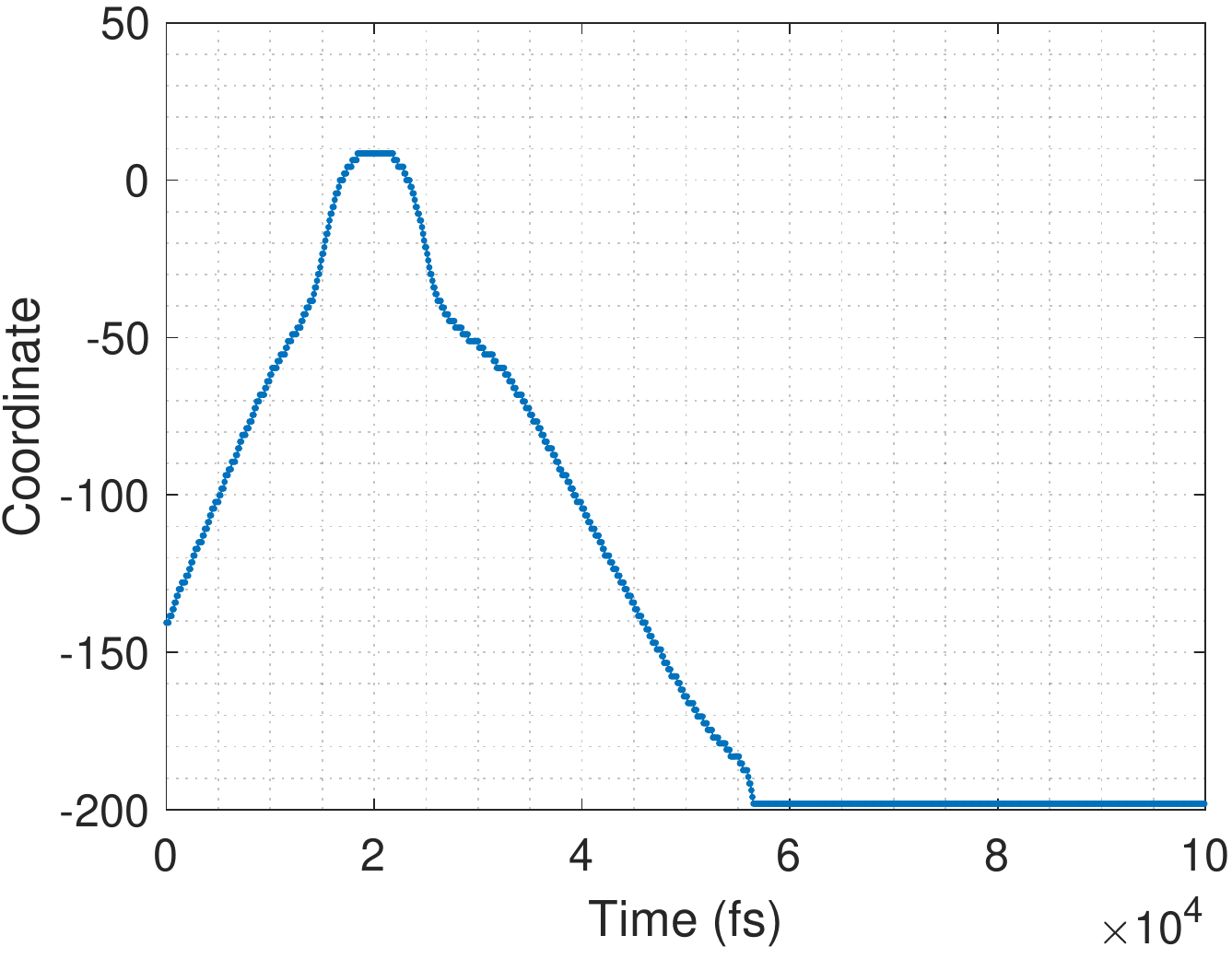}
    \includegraphics[width=0.49\columnwidth]{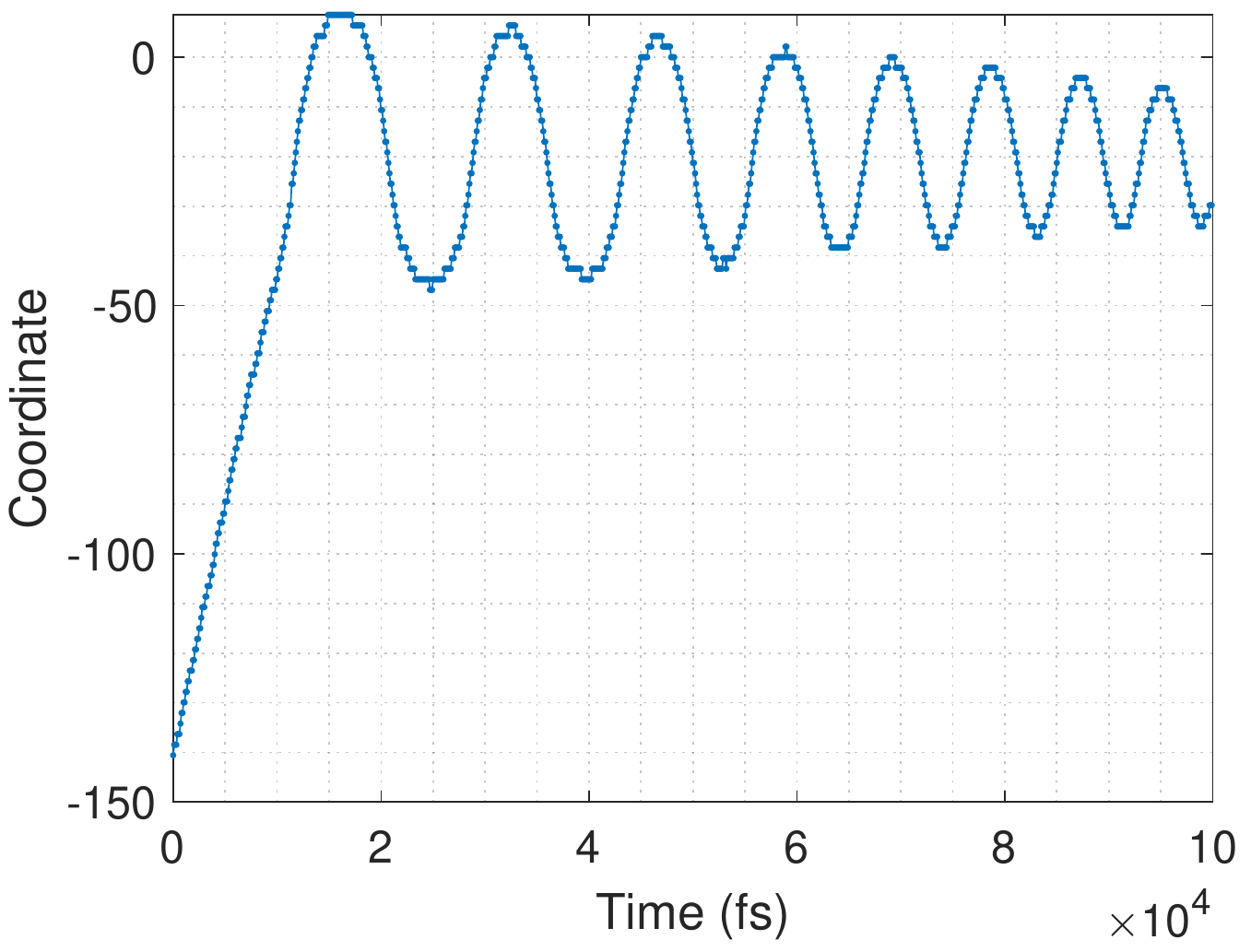}
    \includegraphics[width=0.49\columnwidth]{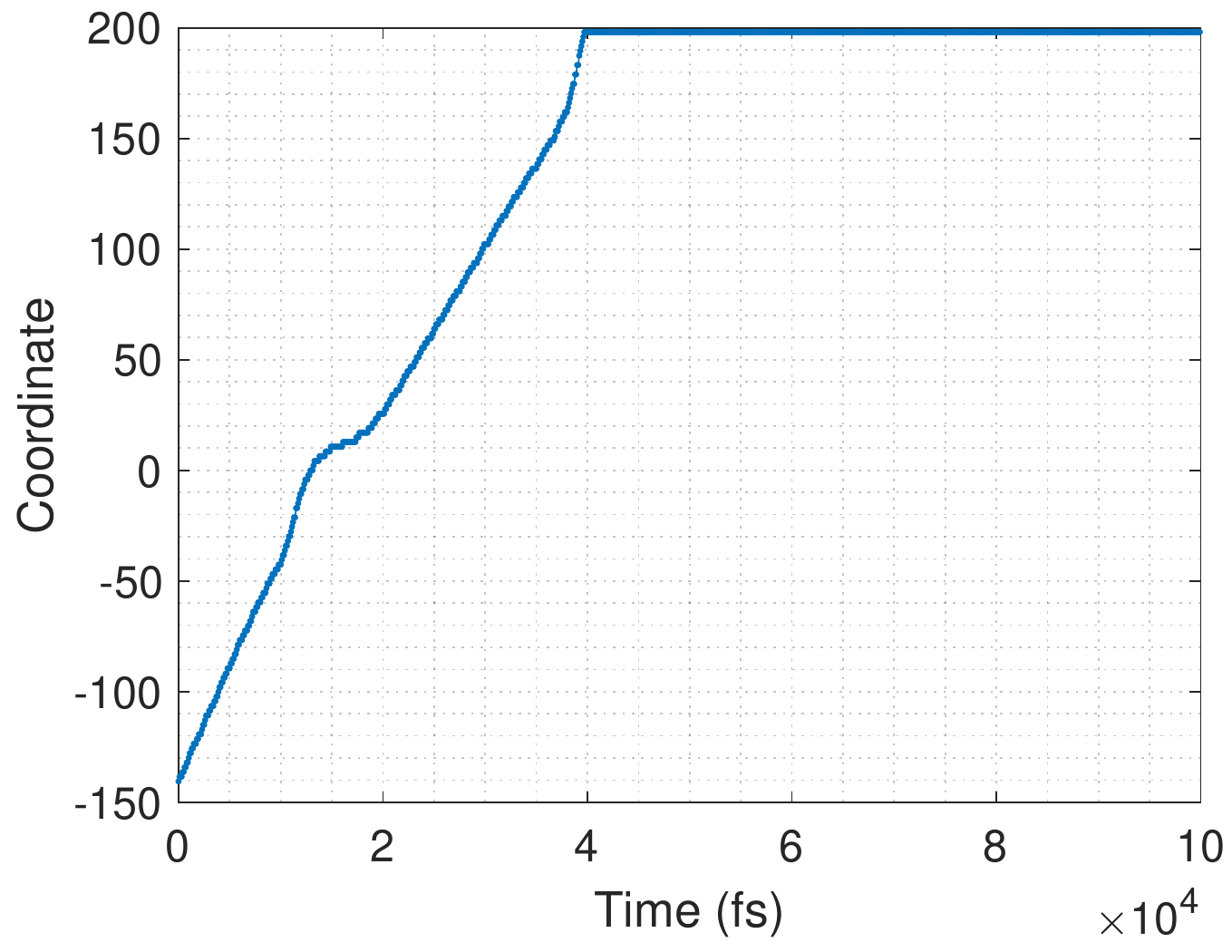}
    \caption{\label{fig:3} Kink scattering by type 1 constriction: Potential energy as a function of $x$ (blue line), and the final position of the kink for different initial velocities (stars). All stars correspond to ($\alpha,+$)-kinks. The coordinate is given in {\AA}. Middle row: Examples of the trajectory for $v_0=0.69$~km/s (left) and $v_0=0.836$~km/s (right). Bottom row: Examples of the trajectory for $v_0=0.988$~km/s (left) and $v_0=1.05$~km/s (right).}
\end{figure}

\subsection{Scattering by symmetric constrictions} \label{sec:3b}

\begin{figure}[tb]
    \centering
    \includegraphics[width=0.9\columnwidth]{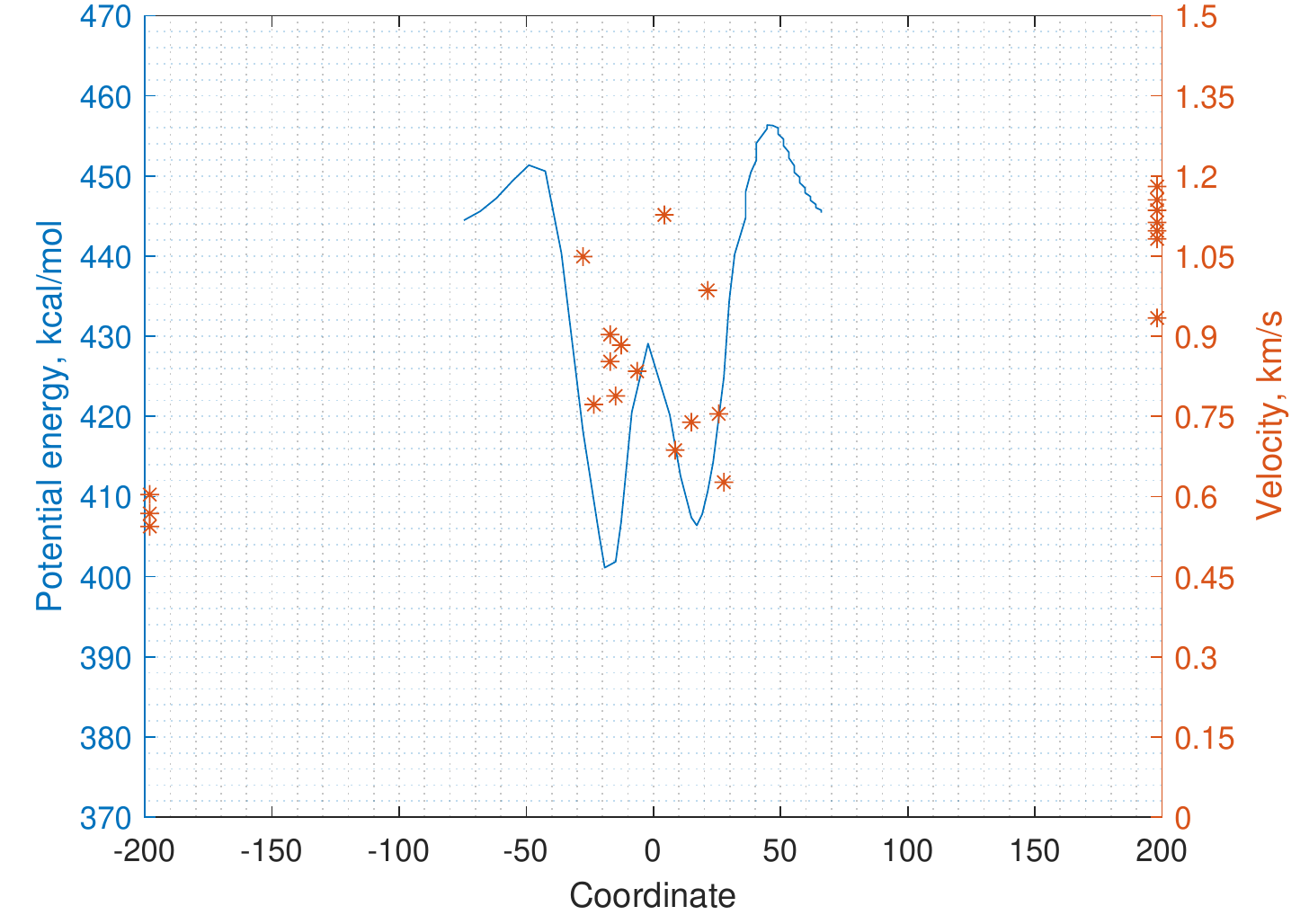}
    \includegraphics[width=0.49\columnwidth]{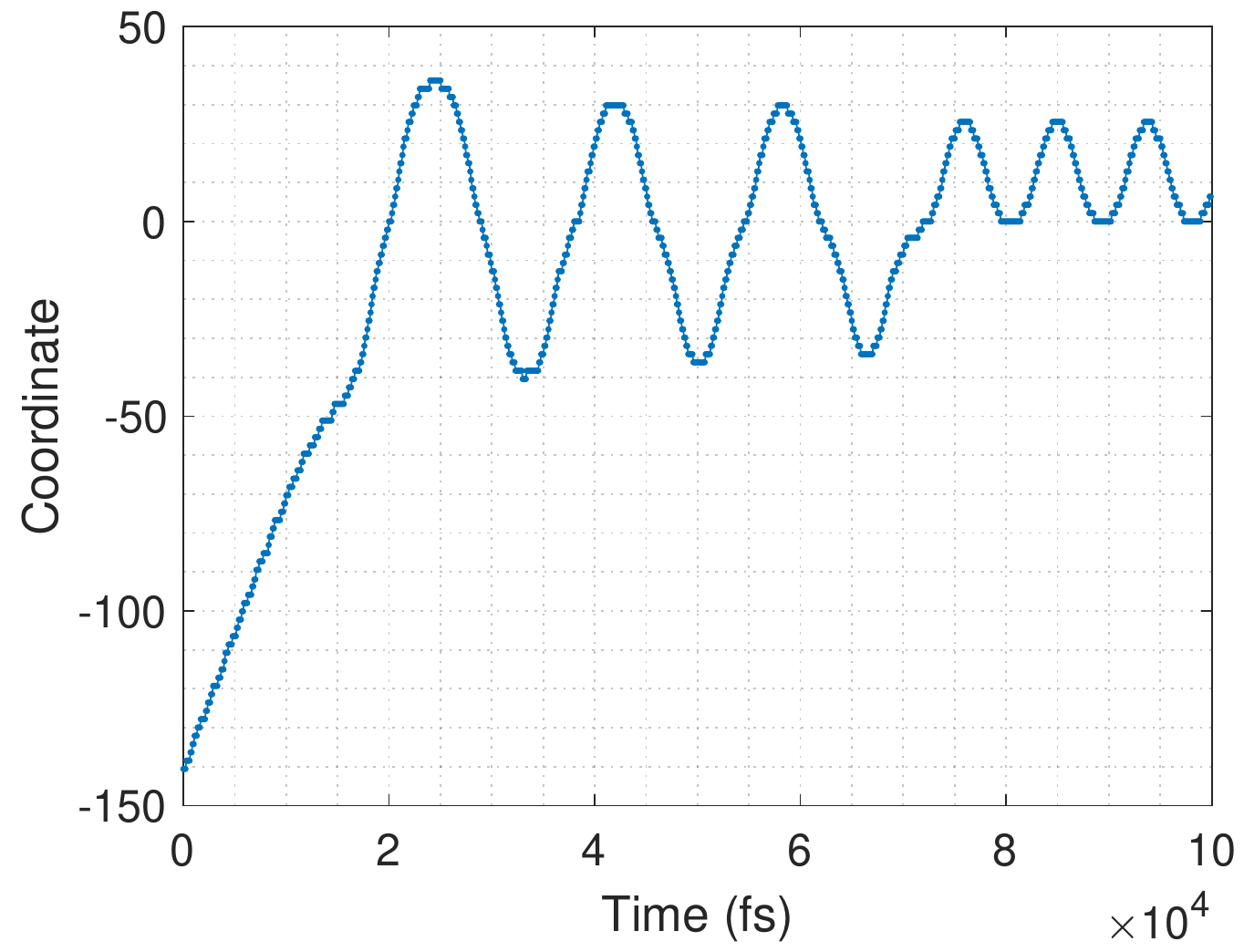}
    \includegraphics[width=0.49\columnwidth]{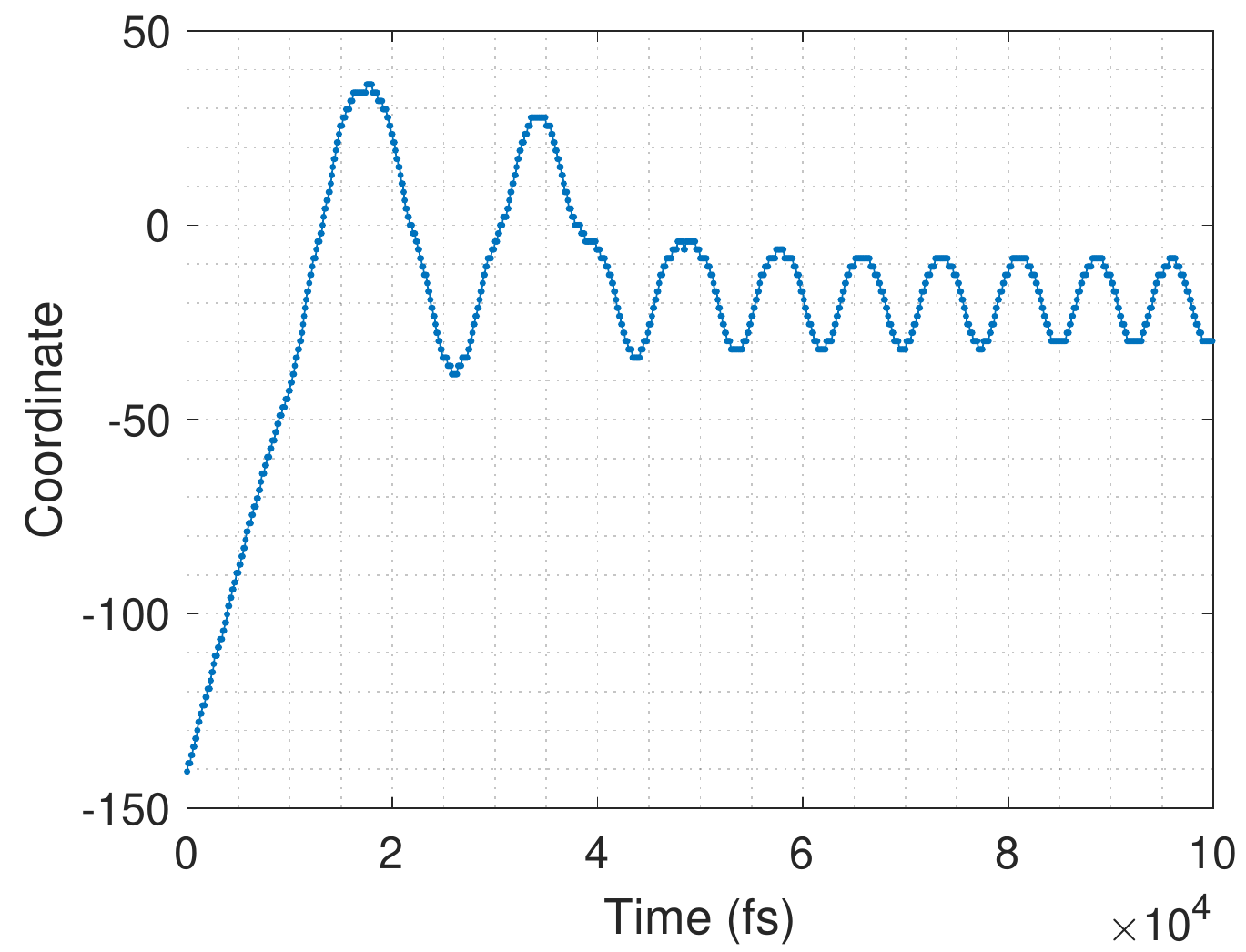}
    \caption{\label{fig:4}Kink scattering by a type 2 constriction. Top: Potential energy as a function of $x$ (blue line), and the final position of the kink for different initial velocities (stars). Here, all stars correspond to ($\alpha,+$)-kinks. The coordinate is given in {\AA}. Bottom row: Examples of the trajectory for $v_0=0.7$~km/s (left) and $v_0=1.05$~km/s (right).}
\end{figure}

Figs.~\ref{fig:3} and \ref{fig:4} present MD simulation results for type 1 and 2 constrictions, respectively. In the top graphs, the orange stars represent the $x$-coordinate of the kink at $t=100$~ps ($10^5$ simulation steps) for different initial velocities. The blue lines show the potential energy of the nanoribbon atoms as a function of the position of the kink. The potential energy was calculated as described in the last paragraph in Sec.~\ref{sec:2}.  The potential energy is shown relative to the energy of the buckled up nanoribbon, which is 8867~kcal/mol. In the middle and bottom graphs in Figs.~\ref{fig:3} and \ref{fig:4} we show examples of the trajectory of the kink. Each example corresponds to a star in the top graphs in Figs.~\ref{fig:3} and \ref{fig:4}.

In classical mechanics, the space dependence of the potential energy is very useful in understanding the motion.  The same principle
can be used to explain the general features of kink dynamics, however, the dynamics of kinks is more complex, as kinks are extended objects with internal degrees of freedom that, in particular, may lose their energy in interactions with each other or in the presence of constrictions and/or barriers. Let us  consider the  potential energy of the atoms of the membrane as a function of the position of the kink (solid lines in the top graphs in Figs.~\ref{fig:3} and \ref{fig:4}). In the case of a symmetric constriction (type 1), the potential energy has a minimum surrounded by relatively low energy barriers (Fig.~\ref{fig:3}, top). The minimum is shifted from the position of constriction in a direction defined by the type of kink (to the left from $x=0$ in Fig.~\ref{fig:3}, top). In the case of a double symmetric constriction (type 2), the potential energy can be thought of as a superposition of the potential energies from two type 1 constrictions, see Fig.~\ref{fig:4}, top.

According to Fig.~\ref{fig:3}, top and a visual analysis of the MD trajectories (for trajectories, see the middle and bottom rows in Fig.~\ref{fig:3}), slow kinks ($v_0 < 0.55$~km/s) are always reflected from the first barrier (formed to the left from the constriction). Slightly faster kinks are reflected from the second barrier (which is higher than the first one), and get trapped by the potential well (Fig.~\ref{fig:3}, middle (left)). Their subsequent dynamics is reduced to dissipative oscillations (some motion was observed after $10^5$ simulation steps). We observed that kinks with $v_0 = 0.795$~km/s and $v_0 = 0.836$~km/s are reflected from the second barrier, pass above the first one, and return to the left edge of the membrane (Fig.~\ref{fig:3}, middle (right)). Such kind of behavior is known in the literature as a resonance~\cite{PhysRevA.46.5214}, which is explained by a resonant energy exchange between the kinetic energy of the kink, its internal mode, and the impurity mode~\cite{PhysRevA.46.5214}. Kinks with $v_0> 0.84$~km/s mainly pass through the constriction (Fig.~\ref{fig:3}, bottom (right)), although a number of trapping resonances can be observed in Fig.~\ref{fig:3}, top. In particular, Fig.~\ref{fig:3}, bottom (left) shows an example of the trapped kink whose initial velocity exceeds the $0.84$~km/s threshold ($v_0=0.988$~km/s). We note that while some resonances are pretty narrow (such as the one at $v_0= 0.988$~km/s represented by a single star in Fig.~\ref{fig:3}, top), some others have a finite width on the scale of Fig.~\ref{fig:3}, top (such as the one at $v_0\approx 1.02$~km/s represented by three stars).

Quite similar behavior is observed in the case of a type~2 constriction, see Fig.~\ref{fig:4}, top. We note that in the top graph in Fig.~\ref{fig:4} the first potential barrier is approximately 60~kcal/mol higher than that in Fig.~\ref{fig:3}, top. Correspondingly, the initial velocity of the kinks passing the first barrier is also higher (it starts at about $0.6$~km/s).
The general picture is the following: reflection of slow moving kinks, trapping of intermediate-velocity kinks, and transmission of fast kinks. In particular, to pass the constriction, the kink's velocity has to be at least 1.07 km/s. When trapped,  kinks oscillate between the two wells, slowly losing their velocity and, eventually, are trapped in one of the wells (see Fig.~\ref{fig:4}, bottom graphs). Moreover, a few cases of resonant behavior can be distinguished in Fig.~\ref{fig:4}, top.

\begin{figure}[tb]
    \centering
    \includegraphics[width=0.8\columnwidth]{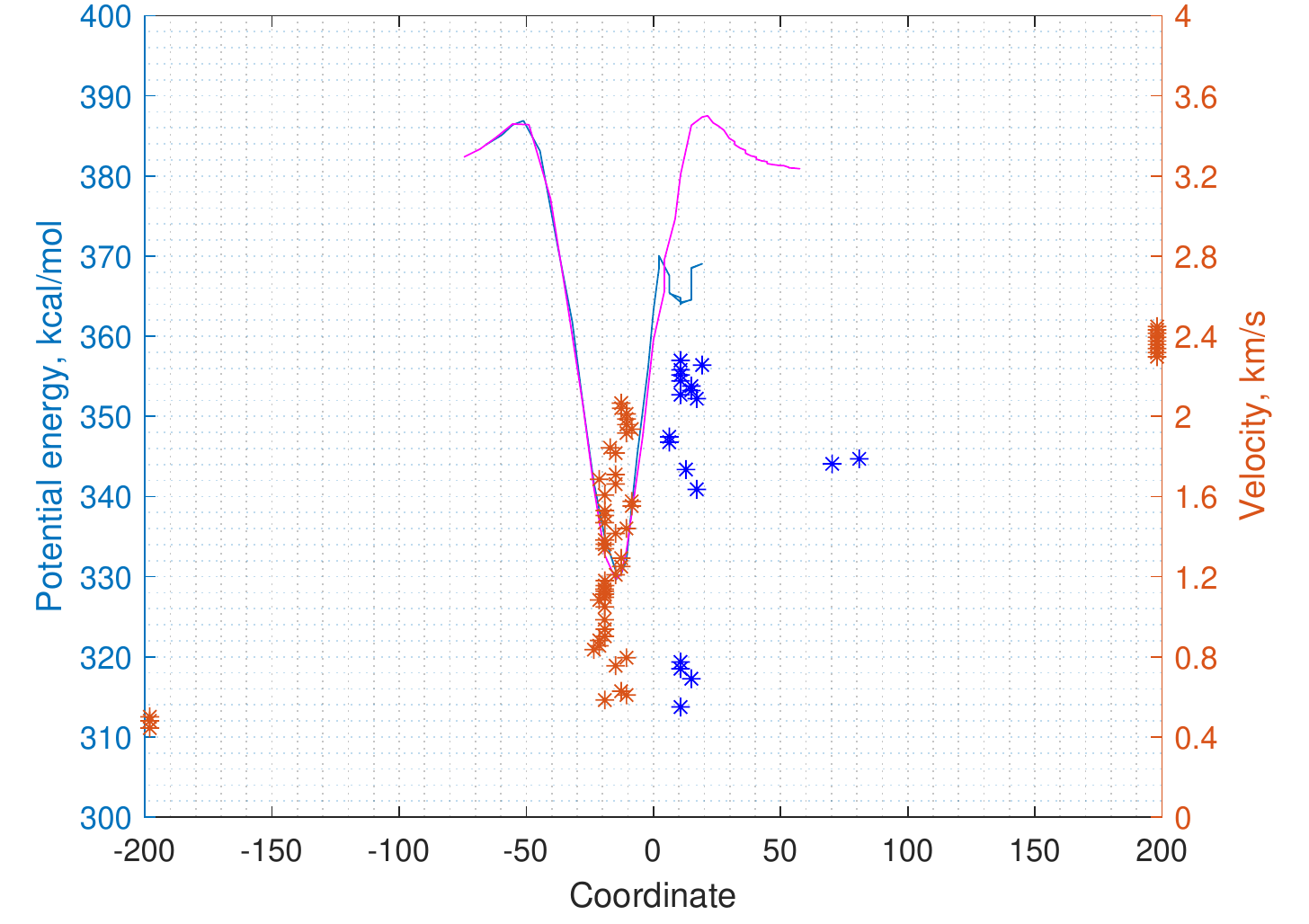}
    \includegraphics[width=0.49\columnwidth]{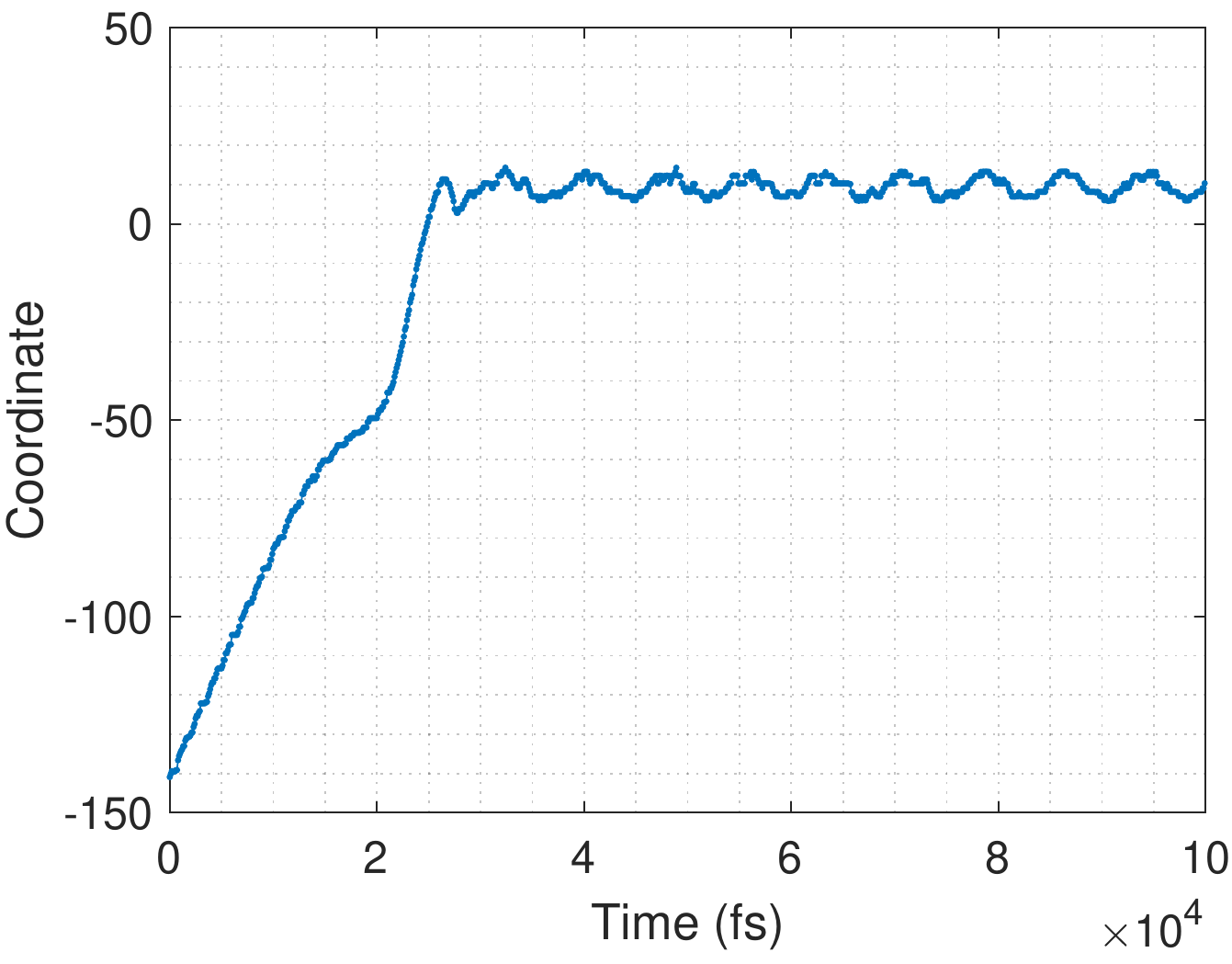}
    \includegraphics[width=0.49\columnwidth]{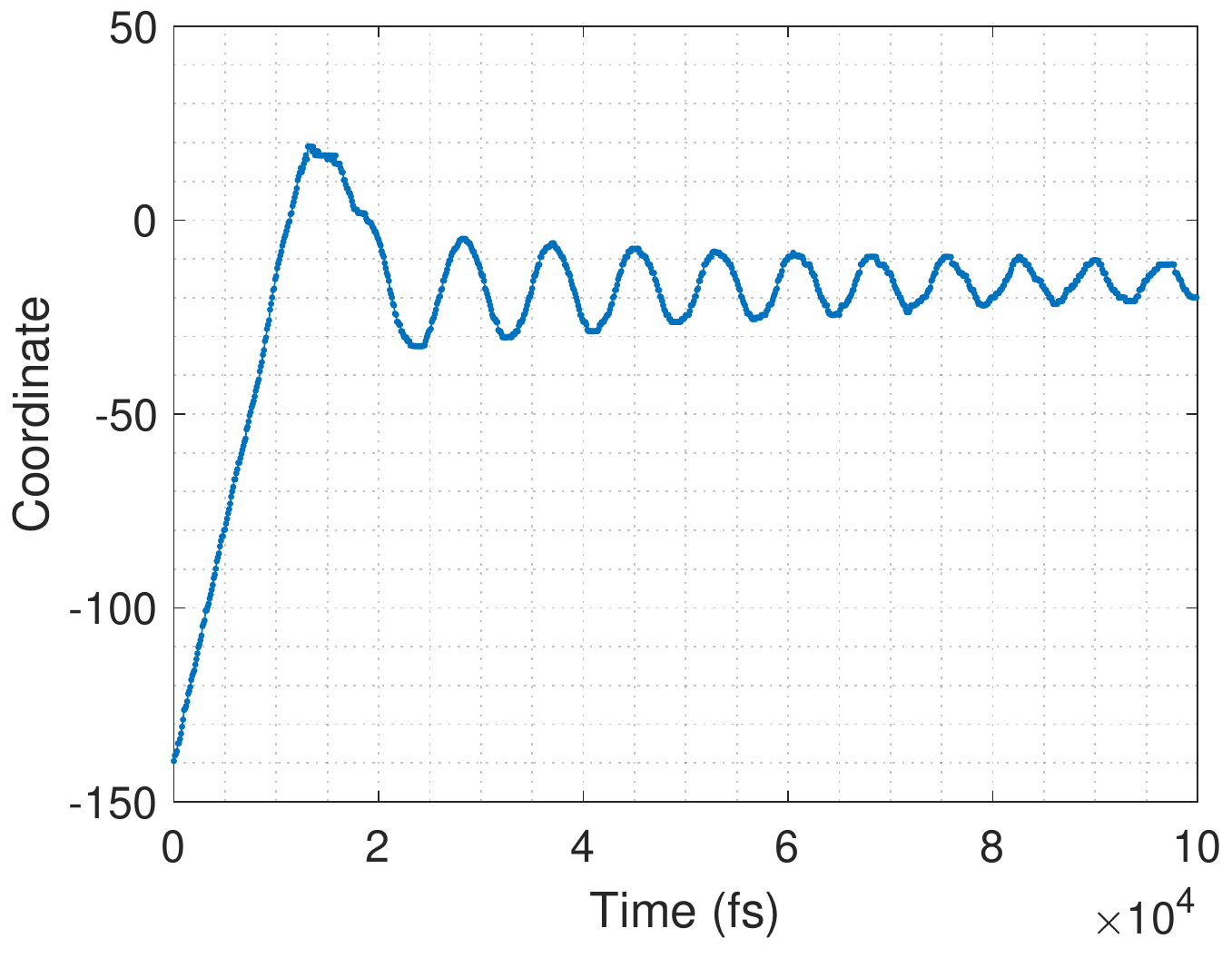}
    \includegraphics[width=0.49\columnwidth]{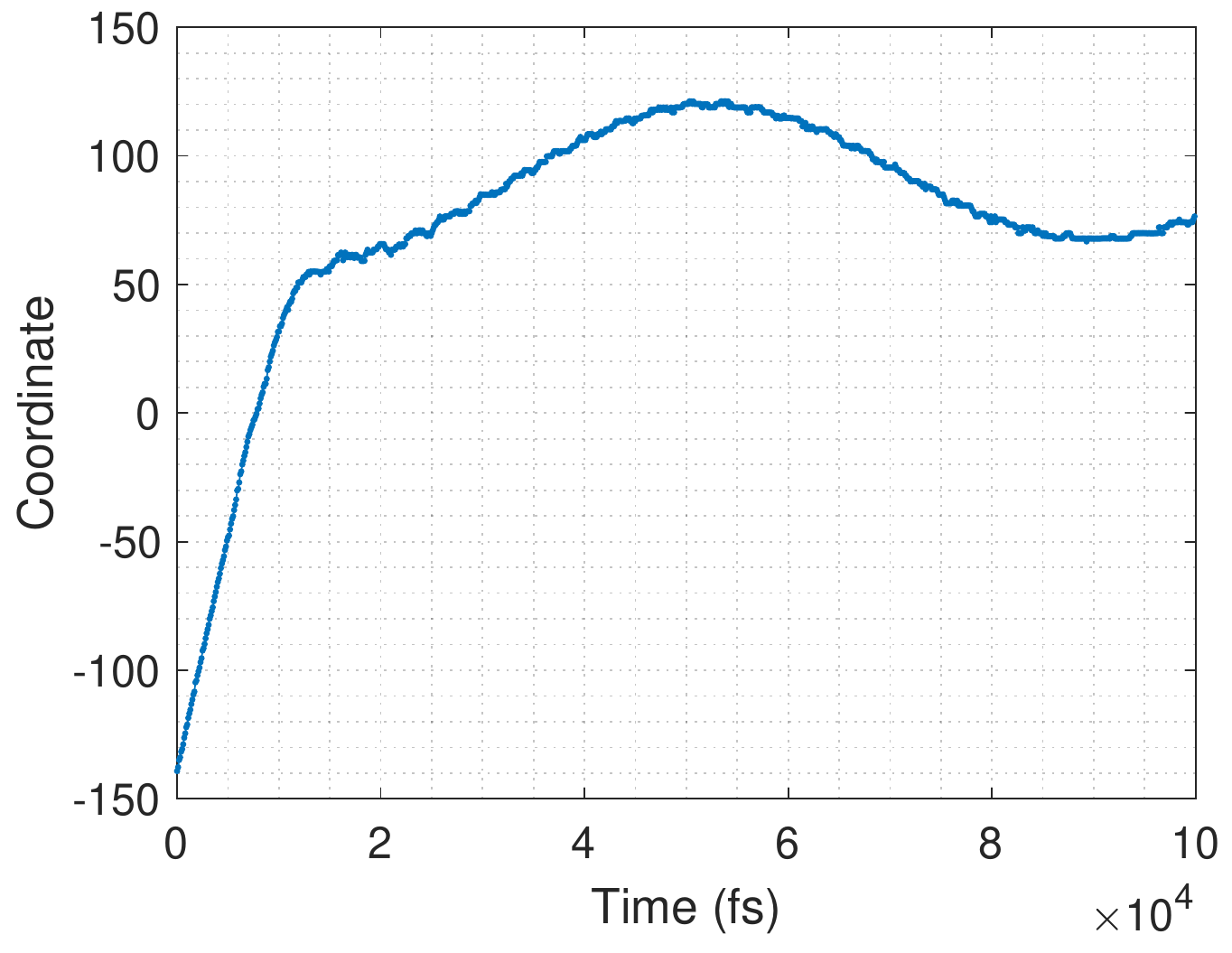}
    \includegraphics[width=0.49\columnwidth]{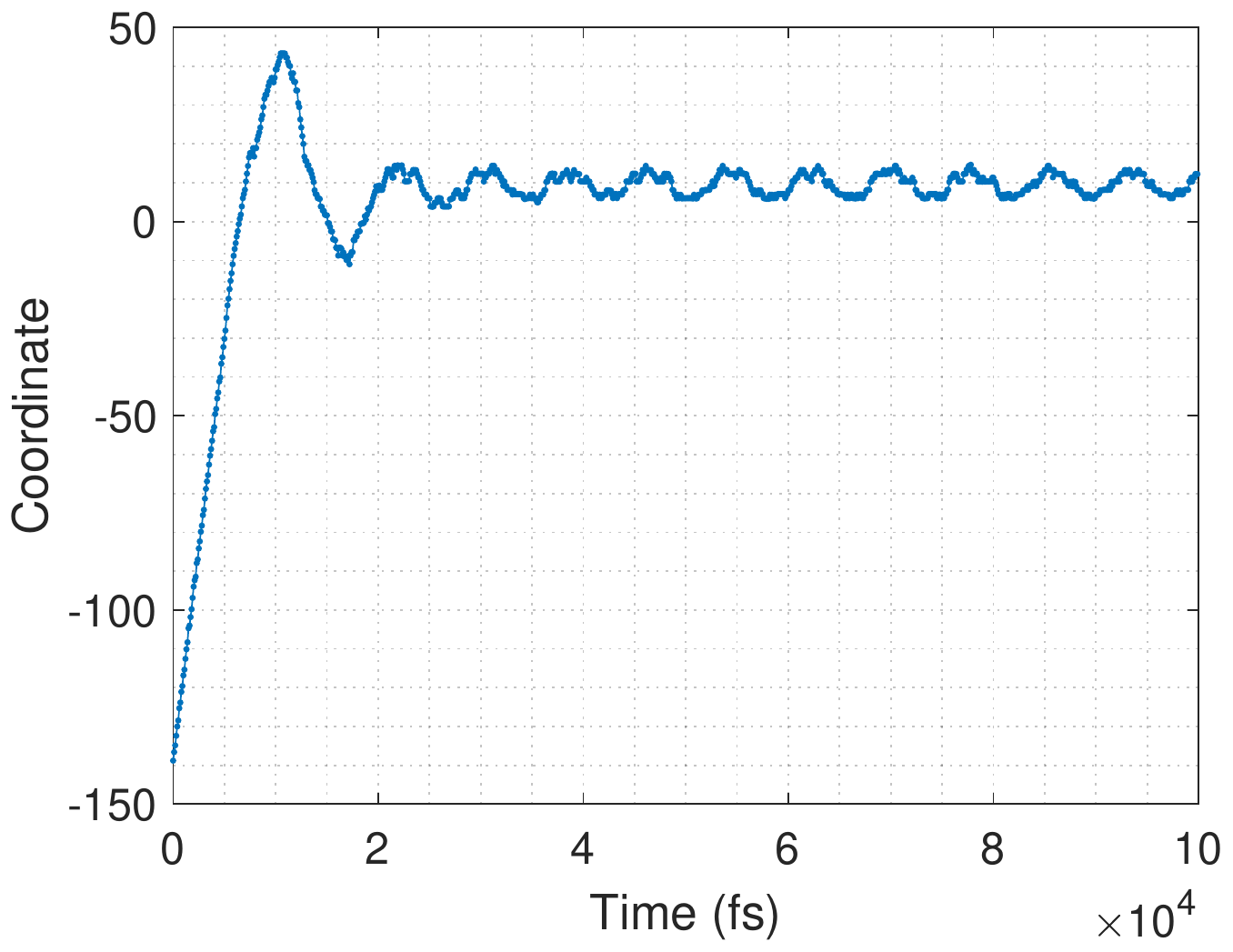}
    \caption{\label{fig:5}Kink scattering by a type 3 constriction. Top: Potential energy as a function of $x$ (magenta and blue lines), and the final position of the kink for different initial velocities (stars). Here, the orange stars correspond to ($\alpha,+$)-kinks, while blue stars represent ($\beta,-$)-kinks. The magenta line was obtained using a fast moving kink; the blue line was obtained using a slow moving kink. The coordinate is given in {\AA}.  Middle row: Examples of the trajectory for $v_0=0.55$~km/s (left) and $v_0=1.15$~km/s (right). Bottom row: Examples of the trajectory for $v_0=1.76$~km/s (left) and $v_0=2.12$~km/s (right).}
\end{figure}

\subsection{Scattering by asymmetric constrictions} \label{sec:3c}

The potential energy of kinks in a membrane with a type 3 asymmetric constriction is shown in the top graph in Fig.~\ref{fig:5}. While this potential energy
resembles the one in Fig.~\ref{fig:3}, top, we have observed a richer and less predictable behavior of the scattering of the the kinks on asymmetric constrictions. The most interesting feature is the possibility of the transformation of ($\alpha,+$)-kinks into ($\beta,-$)-ones.

The magenta line in Fig.~\ref{fig:5}, top, represents the potential energy when the kink does not change its type.  The blue line is for the case when  an $(\alpha,+)$-kink transforms into a $(\beta,-)$-kink. To penetrate inside the well, the velocity of the kink must exceed $0.5$~km/s. At $v_0 = 0.54$~km/s, an incoming ($\alpha,+$)-kink reflects from the second barrier, almost completely stops at the position of the first  barrier, and then accelerates to the right, undergoing a transformation into a ($\beta,-$)-kink, which stops immediately after the transformation. Faster kinks with $v_0 \sim 0.6$~km/s behave similarly to the kinks trapped by type 1 constrictions. At $v_0 = 0.7$~km/s, an ($\alpha,+$)-kink is trapped, performs several oscillations in the well, losing its kinetic energy,  and then transforms into a ($\beta,-$)-kink. Starting at $v_0 = 0.77$~m/s, an incoming ($\alpha,+$)-kink transforms into a ($\beta,-$)-kink, almost immediately transforms back, and oscillates, being trapped by the well. Higher velocity kinks can transform a second time into  ($\beta,+$)-kinks. At $v_0=1.8$~km/s, instead of being trapped or reflected, a ($\beta,-$)-kink first moves forward, then stops, and then moves backwards. Only when the initial velocity is higher than 2.3 km/s does an incoming ($\alpha,+$)-kink transform into a moving ($\beta,-$)-kink, which transforms back into an ($\alpha,+$)-kink to the right of the constriction.

The middle and bottom rows in Fig.~\ref{fig:5} present examples of the trajectory for several selected values of $v_0$. We found that in these graphs, the ($\beta,-$)-shape of the kink is characterized by a higher level of noise than the ($\alpha,+$)-shape. According to this observation, in all but the middle (right) graph in Fig.~\ref{fig:5}, the final shape of the kink is ($\beta,-$).

\begin{figure}[tb]
   \centering
   \includegraphics[width=0.75\columnwidth]{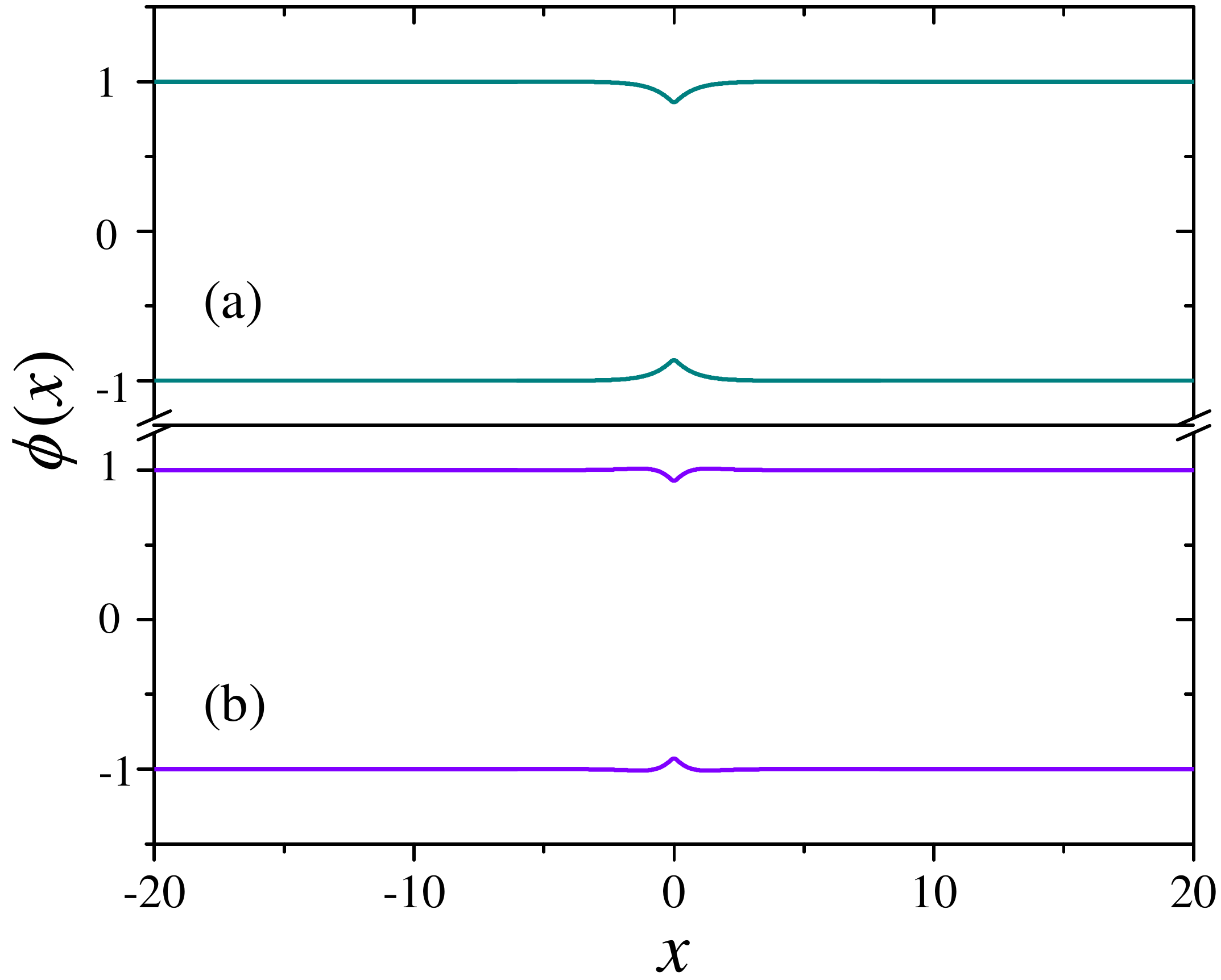}
    \caption{\label{fig:model} Equilibrium solutions of Eq.~(\ref{eq:2aa}) with (a) $\gamma(x)=G_{\sigma_1}(x)$, and (b) $\gamma(x)=G_{\sigma_1}(x)-kG_{\sigma_2}(x)$,
where  $G_\sigma(x)$ is a Gaussian function. We used the following parameter values: $\epsilon=0.45$, $k=8/9$, $\sigma_1=0.08$,  $\sigma_2=0.8$, $x_0=0$. }
\end{figure}

\subsection{$\phi^4$ model with constriction} \label{sec:3d}

In this subsection we introduce a constriction into the classical $\phi^4$ model and perform simulations of $\phi^4$ kink dynamics.
Previously, the $\phi^4$ model with impurities has been studied by several authors~\cite{PhysRevA.46.5214,kink_and_realistic_impurity}. The impurity potential
was specified by the following substitution in Eq.~(\ref{eq:1}):
\begin{equation}\label{eq:imp1}
   \frac{1}{4}(1-\phi^2)^2\rightarrow \frac{1}{4}(1-\phi^2)^2\left(1-\epsilon\gamma(x-x_0) \right),
\end{equation}
where $\epsilon$ is the impurity strength, $\gamma(..)$ is a function describing the shape (e.g.,  Gaussian or Lorentzian~\cite{kink_and_realistic_impurity}), and $x_0$ is the location of the impurity.
However, while the right-hand side of~(\ref{eq:imp1}) locally modifies the potential energy, the value of $\phi$ that minimizes the right-hand side of~(\ref{eq:imp1}) is $\pm 1$. Therefore, the above substitution can not describe a local reduction of $\phi$ at the constriction.

Below, we report results of numerical simulations using a slightly different imperfection model,
\begin{equation}\label{eq:imp2}
   \frac{1}{4}(1-\phi^2)^2\rightarrow \frac{1}{4}(1-\epsilon\gamma(x-x_0)-\phi^2)^2
\end{equation}
leading to the equation of motion
\begin{equation}
    \frac{\partial^2 \phi}{\partial t^2} - \frac{\partial^2 \phi}{\partial x^2} + \phi \left(\phi^2 - (1-\epsilon\gamma(x-x_0)) \right) = 0 .
    \label{eq:2aa}
\end{equation}
The equilibrium lowest-energy solutions of Eq.~(\ref{eq:2aa}) are shown in Fig.~\ref{fig:model} for the cases when $\gamma(x-x_0)$ is a Gaussian (Fig.~\ref{fig:model}(a)) and a mixture of two Gaussians (Fig.~\ref{fig:model}(b)). To obtain Fig.~\ref{fig:model}, Eq.~(\ref{eq:2aa}) was solved numerically with an extra dissipation term $-\tau^{-1}\partial \phi / \partial t$ in its left-hand side and the initial condition $\phi(x,t=0)=\pm 1$, using the Euler integration method (with $\tau=1$).

\begin{figure*}[tb]
 (a) \includegraphics[width=0.6\columnwidth]{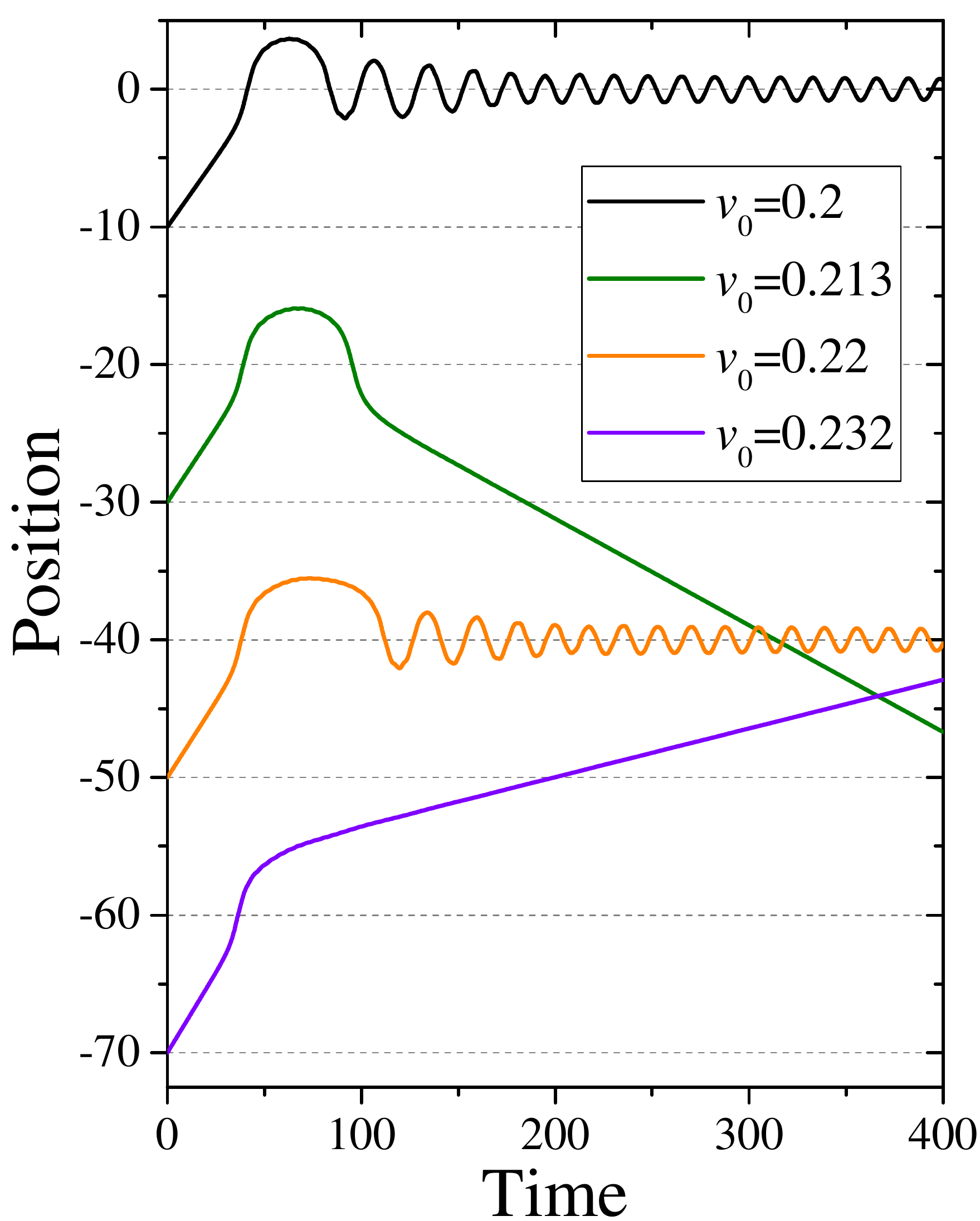} \;\;\;\;\;
  (b)\includegraphics[width=0.6\columnwidth]{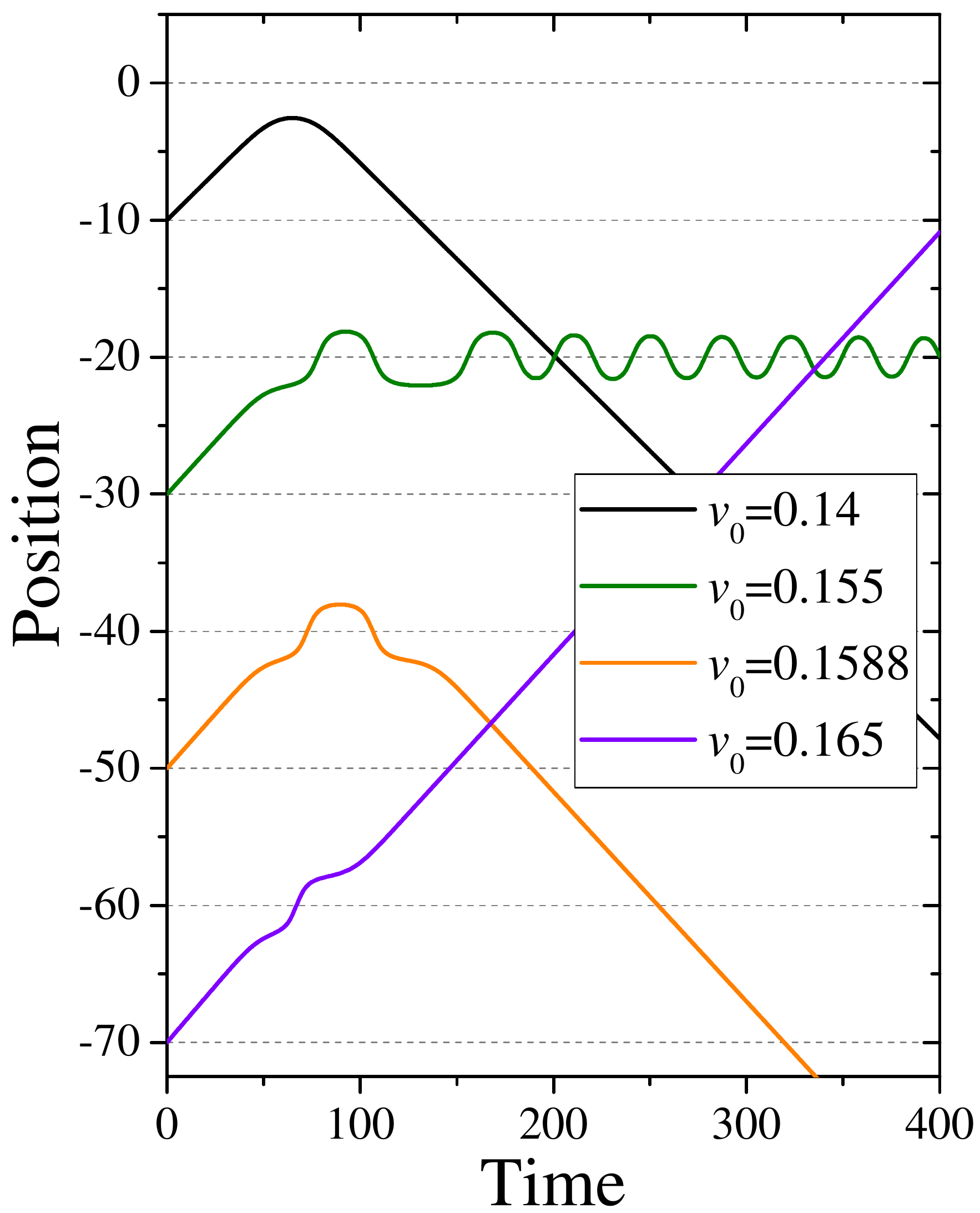}
    \caption{\label{fig:7} Position of the kink as a function of time for different initial velocities, $v_0$, obtained using (a) $\gamma(x)=G_{\sigma_1}(x)$ and (b)  $\gamma(x)=G_{\sigma_1}(x)-kG_{\sigma_2}(x)$ in Eq.~(\ref{eq:2aa}) with parameters as in Fig.~\ref{fig:model}. The curves were displaced vertically for better viewing.}
\end{figure*}

The initial conditions for the simulations of kink scattering (Fig.~\ref{fig:7}) were obtained by mixing the $\phi(x)>0$ solution in Fig.~\ref{fig:model} with the kink profile, Eq.~(\ref{eq:3}). The initial position of the kink was set at $a=-10$. Eq.~(\ref{eq:2aa}) was solved numerically with a time step of $0.01$ and the space discretization step of $0.02$.

Fig.~\ref{fig:7}(a) shows examples of the kink's trajectory found with a single Gaussian model of the constriction,  $\gamma(x)=G_{\sigma_1}(x)$, in Eq.~(\ref{eq:2aa}). We found that for initial velocities below the threshold $v_t\approx 0.23$, the kink is trapped by the constriction except for several narrow resonances wherein the kink is reflected back. An example of resonant reflection is the graph for $v_0=0.213$ in Fig.~\ref{fig:7}(a). Kinks with initial velocities above the threshold propagate through the constriction.

\begin{figure}[tb]
    \includegraphics[width=0.75\columnwidth]{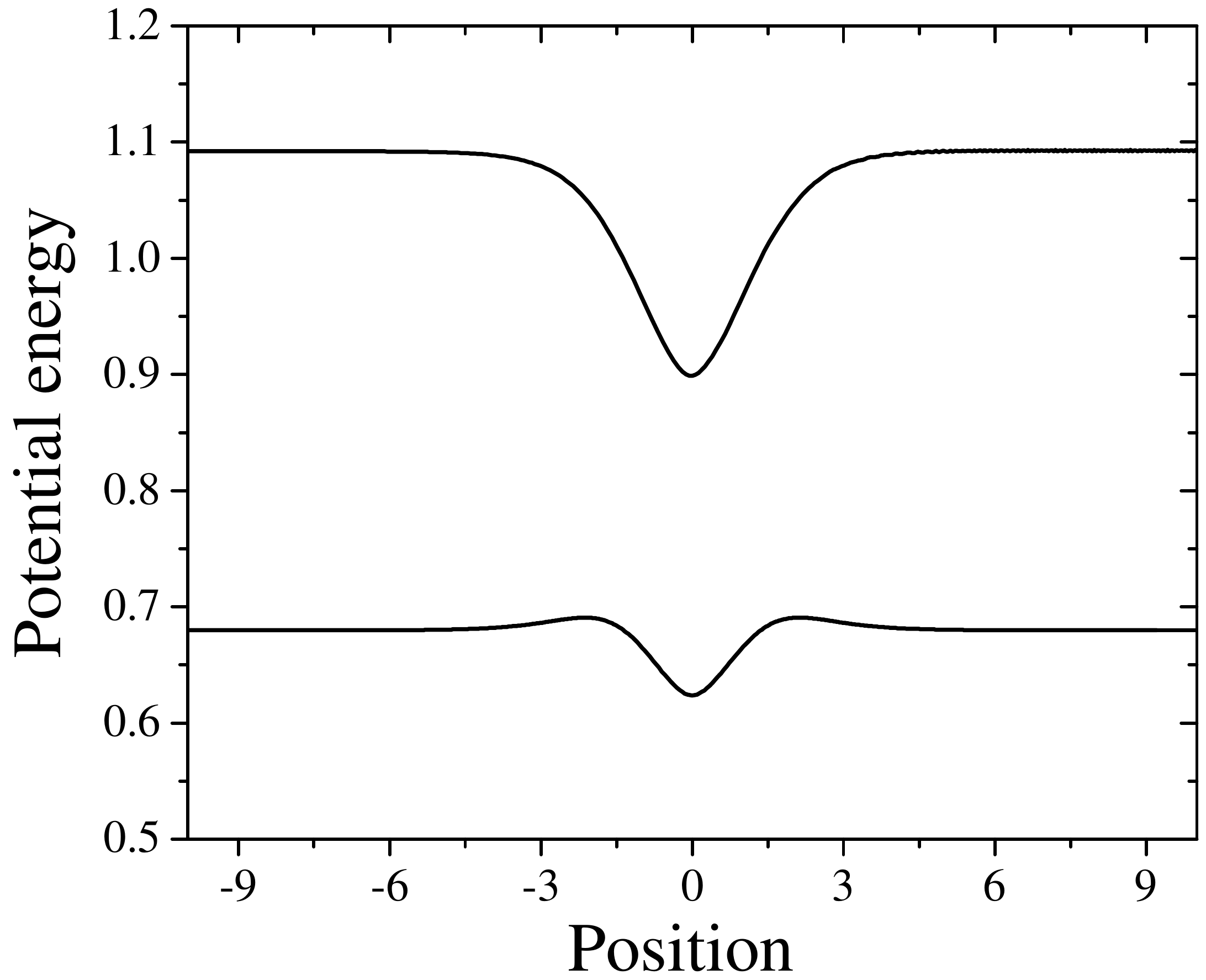}
    \caption{\label{fig:8}Potential energy as a function of the position of the kink, obtained using $\gamma(x)=G_{\sigma_1}(x)$ (top curve) and  $\gamma(x)=G_{\sigma_1}(x)-kG_{\sigma_2}(x)$ (bottom curve) with parameters as in Fig.~\ref{fig:model}. The bottom curve was displaced down by 0.4 for clarity.}
\end{figure}

Clearly, the $\gamma(x)=G_{\sigma_1}(x)$ model of constriction can only reproduce the attractive part of the constriction potential experienced by graphene kinks. To introduce the repulsive barriers, we use a double Gaussian model, $\gamma(x)=G_{\sigma_1}(x)-kG_{\sigma_2}(x)$.
Selected simulation results found with the double Gaussian model are presented in Fig.~\ref{fig:7}(b). The behavior of the kink is characterized by two well-defined thresholds, $v_{t,1}\approx 0.15088$ and $v_{t,2}\approx 0.16155$. Kinks are always reflected for $v_0<v_{t,1}$, and transmitted for $v_0>v_{t,2}$. For intermediate values of $v_0$, except for several narrow resonances (such as $v_0=0.1588$ in Fig.~\ref{fig:7}(b)),  the kinks are trapped by the constriction.

Fig.~\ref{fig:8} shows the potential energy as a function of the position of the kink for the cases of single and double Gaussian models. The potential energy was calculated using the same approach as we used in the calculations of the potential energy of graphene kinks. We ran a series of similar simulations of kink scattering (as shown in Fig.~(\ref{fig:7})) with the same initial velocity (selected above the transmission threshold) but different number of simulations steps. Each step was followed by minimization. For this purpose, the velocities of all field elements were set to zero, and Eq.~(\ref{eq:2aa}) was simulated for 300 time steps with an extra dissipation term $-\tau^{-1}\partial \phi / \partial t$ in its left-hand side with $\tau=1$. Fig.~\ref{fig:8} demonstrates that the double Gaussian model makes a closer approximation to the potential experienced by the graphene kinks in the membrane with type 1 constriction (Fig.~\ref{fig:3}, top).

\section{Conclusion}\label{sec:4}

Transport through barriers and constrictions have been of significant importance in various fields such as quantum mechanics~\cite{razavy2003quantum,landau2013quantum}, solid-state physics~\cite{di2008electrical}, and bio-nano techology~\cite{branton2010potential}. In this paper, the dynamics of a kink in  buckled graphene with  constriction has been investigated using the method of molecular dynamics simulations.
It has been observed that the interaction of a graphene kink with a constriction is complex, and the main component in the interaction is attractive. The simplest symmetric constriction can be considered as a potential well surrounded by two relatively low repulsive barriers. In some of our calculations, we have observed scattering resonances probably similar to the ones in  the $\phi^4$ model with an impurity~\cite{PhysRevA.46.5214,kink_and_realistic_impurity}.

A rich dynamical behavior has been observed in the case of asymmetric constrictions, including the possibility of the transformation of the type of the kink. This type of constriction can be used as a tool to create $\beta$-type kinks in numerical or physical experiments. Finally, we have introduced a constriction potential into the $\phi^4$ Lagrangian, and simulated the $\phi^4$ kink dynamics in the presence of constriction. The behavior is qualitatively similar to that observed in molecular dynamics simulations. It is still an open question of how to describe asymmetric constrictions (and several kink types) in the framework of $\phi^4$ theory. This may be the subject of future research.

RDY acknowledges support from RFBR grant 19-32-60012. The force field used in this study as well as the data supporting our findings are available from the corresponding author upon reasonable request.

\bibliographystyle{apsrev4-1}
\bibliography{bibliography}

\end{document}